\documentclass[12pt]{iopart}

\usepackage{graphicx}
\usepackage[bookmarks=true,colorlinks=true,linktocpage=true,backref=true]{hyperref}
\usepackage{endnotes}
\usepackage{hyperendnotes}
\usepackage{amsfonts}
\usepackage[table]{xcolor}

\begin{document}

\bibliographystyle{iopart-num}

\title[Expanding and SOC Emerging from Frozen TSN]{Expanding and Self-Organizing $2$D Universe Models \\
Emerging from Frozen Trivalent Spin Networks}

\author{Christine C. Dantas}
\address{Divis\~ao de Astrof\'{\i}sica (INPE-MCTI), 
Av. dos Astronautas, 1758, Jardim da Granja, 
CEP: 12227-010, S\~ao Jos\'e dos Campos, SP, Brazil }
\ead{christine.dantas@inpe.br}

\date{\today}

\begin{abstract}

We revisit the topic of self-organized criticality (SOC) in simple statistical graph models, with the purpose of capturing essential processes leading to the emergence of macroscopic spacetime from the microscopic dynamics in loop quantum gravity (LQG). We performed a large set of simulations based on extensions of the frozen trivalent spin network (TSN) model explored previously by Ansari and Smolin. Their model mimicked the sandpile dynamics by the application of random vertex propagation rules in the TSN, leading to a SOC behavior in the distribution of the avalanche sizes, as well as a slowly expanding, $2$-dimensional dual (triangulated) space. Here we show that a growth scheme for the stochastic, slow external driving force, differing from the classical sandpile model, also resulted in power-law distributed avalanche sizes. Our simulations also produced expanding dual spaces, with two basic classes of evolution: one with power-law correlations in ``space'' and ``time'', and the other with ``loitering'' and exponential phases. Our work expands the range of models in which critical states in the TSN may lead to expansion effects in the dual space, without fine-tuning.

\end{abstract}


\maketitle


\section{Introduction \label{INTR}}

The expectation that physical interactions should be fundamentally unified around the Planck scale ($\ell_{\rm Pl} \equiv \sqrt{G\hbar/c^3} \sim 10^{-33} ~{\rm cm}$), as well as the presence of singularities in general relativity, suggest that gravity should be quantized. Although a theory of quantum gravity has not been established yet, there has been robust developments in the past decades, based on distinct motivations and approaches. These can be roughly traced back from two fronts \cite{Rov98}:  the particle physics community (unification of interactions), and the general relativity community (breakdown of the smooth, classical spacetime). In the latter front, developments like Loop Quantum Gravity (LQG) \cite{RovBOOK07} \cite{ThiBOOK08} \cite{RovVidBOOK20} incorporate a relational, background independent framework at its foundation \cite{Smo05}, leading to the prospect of a non-perturbative canonical quantization of general relativity.

Abstract graphs were first envisioned by R. Penrose as a possible model of spacetime geometry \cite{Pen71}. Graphs are represented in LQG in terms of spin networks, roughly defined as follows \cite{Smo04}, \cite{RovVidBOOK20}. A spin network state is expressed in terms of a graph, $\gamma$, embedded in a spatial manifold $\Sigma$; the graph is composed of a finite number of vertices (or nodes) and incident edges. The edges are labelled by an irreducible representation of SU($2$), namely, spins $j_k$, whereas vertices are labelled by  intertwining operators (based on the tensor product of the representations carried by the incoming and outgoing edges). The physical interpretation is that each node inside an elementary region of $\Sigma$ contributes a quanta of volume, whereas each edge crossing the corresponding surface of the region contributes a quanta of area, being therefore quantum states of space. 

Spin network states form a basis, at the kinematical level, for the representative quantum state functions of $3$-dim space. Physical states would arise from the solution of the Hamiltonian constraints, which generate the dynamics \cite{ThiBOOK08}. Quantum dynamics is one of the main unsolved problems in LQG. One possibility was based on the concept of ``sum-over-histories", inspired by Feynman's functional integral quantization scheme, the so-called ``spin-foam models" \cite{Bae00}. A complete knowledge of the evolution of gravitational  quantum states as well as the  retrieval of Einstein's equations in the continuum limit of LQG are the main open problems of the theory \cite{Rov98},\cite{RovBOOK07},\cite{RovVidBOOK20}. 

One intriguing idea, proposed initially by Markopoulou and Smolin \cite{Mar97}, was the possibility that the underlying quantum spacetime could present critical behavior, analogous to a self-organized critical (SOC) phenomena \cite{Bak87}, \cite{Bak88}, \cite{Dha90}, \cite{Marko14}, \cite{Wat16}, resulting in the possible emergence of classical spacetime from such a process. The SOC concept was introduced by Bak et al. in 1987 \cite{Bak87}, in order to address systems displaying power law behavior (scale invariance), without requiring fine-tuning parameters to achieve such a state. It can be illustrated with the simplest model (the sandpile model), in which a sand grain is dropped randomly on a sandpile, and subsequently it may or may not provoke a downward avalanche of grains, depending on some critical condition. The avalanche terminates whenever the sandpile reaches an new equilibrium condition. At each time a sand grain is dropped, the resulting avalanche, if it occurs at all, may be of a different size (i.e., given by the number of grains involved in the avalanche, noting that any given grain may or may not get repeatedly involved in the avalanche). Therefore, the evolution of the system, triggered by external, stochastic disturbances, is such that it spontaneously moves towards critical states. Hence, ``avalanches'' generically refer to multiscale cascades of state changes in SOC systems, leading to ongoing reconfigurations in such systems. Note that avalanches may transport energy (or any other relevant physical quantity, which is being affected) to arbitrary large scales in the system.

If the discrete structure of spacetime in the Planck regime is a SOC system, then it would be an interesting possibility that the classical spacetime limit would require no fine-tuning to be achieved. Note also that this idea arises as an alternative approach to the problem of the action of the Hamiltonian constraint in LQG \cite{Rov98}, \cite{RovBOOK07}, \cite{RovVidBOOK20}, in which one looks for consistent rules for the evolution of the kinematic states leading to a possible recovery of the action of that constraint in the classical limit. 
That idea was investigated by Borissov and Gupta \cite{Bor99}, in context of the so-called ``frozen'' {\it trivalent spin networks} (TSN). They devised a random edge model with specific propagation rules (avalanches of edge color updates after a perturbation of the system), so that it generated a dynamics in which the system evolved to gauge invariant states after every avalanche. Their results, however, did not produce a critical behavior (i. e., a power-law distribution of avalanches) in the TSN. Subsequently, Ansari and Smolin \cite{Ans08} (hereon [Ans08]) revisited the problem in the same frozen TSN context but using a random vertex model, founding evidence for SOC behavior (see [Ans08] for a detailed explanation on the possible reasons for the failure of reaching a SOC behavior in the random edge model of Borissov and Gubta). Chen and Zhu \cite{Che08} (hereon [Che08]) extended the random edge model of Borissov and Gupta, by including a probability distribution for the edge color propagation, as well as considering a non-fixed value for the color change, founding SOC behaviors for this model, but not in the random vertex model (by the application of their proposed model extension), differing therefore from the results by [Ans08] under simpler conditions (the models of [Ans08] and [Che08] are briefly reviewed in the next section).

The main question addressed in our paper is whether it is possible to obtain expanding universe models at higher rates than in previous works, but still producing SOC behavior. In order to investigate this possibility, we adopted a growth scheme for the externally driving perturbations (therefore differing from a simple sandpile model, in which the perturbation is always set to be a fixed, small scale one). We revisited the model by [Ans08], combining it with more flexible values for the edge color updates, somewhat similarly to [Che08], but here implementing a growth scheme (using $3$ different models), which has not been explored previously.  

Our paper is organized as follows. In Sec. \ref{Sec-TSN}, we review the frozen TSN model, and summarize the main concepts of SOC. In Sec. \ref{Sec-NewTSN}, we introduce the models and simulations, as well as a possible interpretation of dynamics (i.e., time evolution) in these models. In Sec. \ref{Sec-RES}, we present our results, which are summarized and discussed in Sec. \ref{Sec-SUMM}. The \ref{App} presents supplementary results on the simulations executed for the smaller TSN.

\section{The frozen TSN model and SOC \label{Sec-TSN}}

\subsection{Description of the TSN, propagation rules, and previous results. \label{Sec-TSN-1}}

A spin network can be modelled as a labelled graph, $\gamma$, with $N_{\rm v}$ number of vertices (nodes), so that each of its vertices, ${\rm v}_i$ ($i = 1, \dots, N_{\rm v})$, has a certain number of oriented edges $e_1, \dots e_m$. These edges are labelled by the irreducible representation of the SU($2$) group in $(3+1)$-dim LQG, that is, in terms of spins, $j_{\beta}$, or colors, given by $c_{\beta} = 2j_{\beta}$.

Following [Ans08], we study open, $2$-dim trivalent ($m=3$) spin networks (TSN), so that the the dual space to the TSN is a (planar) triangulation of a region of space. The physical length of a side of the triangle is given by $l_{\beta} \sim (1/2) c_{\beta} l_{\rm Planck}$. The gauge-invariance constraint on any given vertex ${\rm v}_i$, with edge colors $(c_1=a,c_2=b,c_3=c)$, is given by:

\begin{equation}
a+b \geq c;  ~~a+c \geq b;  ~~b+c \geq a;
\label{EQN-CONSTR1}
\end{equation}
\begin{equation}
a+b+c ~ = ~{\rm even}.
\label{EQN-CONSTR2}
\end{equation}

In [Ans08], the proposed propagation rules for the evolution of spin network states was given by simultaneously adding or subtracting a fixed color on all edges belonging to a given vertex. Specifically, the algorithm was the following:

\begin{enumerate}
\item{Initialize the TSN by labelling each edge by a random color, so that the gauge conditions (Eqs. \ref{EQN-CONSTR1} and \ref{EQN-CONSTR2}) are satisfied at each vertex in whole TSN. In [Ans08], the initial colors were randomly chosen from integers $\mathsf{C}_{\rm init} = \{10, \dots 20 \}$ (even). }
\item{Choose a random (activating) vertex upon which the external driving disturbance will act, by subtracting a color disturbance, $\Delta c = 2$, from all its edges. If at least one of the edges is $0$, do not act and choose another vertex. }
\item{Test for the gauge conditions (Eqs. \ref{EQN-CONSTR1} and \ref{EQN-CONSTR2}) by sweeping the TSN with a fixed ordering. In [Ans08], this ordering was: left to right and top to bottom.}
\item{For each gauge non-invariant (GNI) vertex, add the color disturbance $\Delta c = 2$ to each of its edges.}
\item{Repeat (iii) until the TSN is completely gauge-invariant again. The {\it size of the avalanche} is given by the number of updates (color additions) imposed on the vertices (repeated vertices are counted). The {\it area of the avalanche} is given by the number of individual vertices involved in the avalanche (not counting how many times they eventually were updated in the same avalanche).}
\item{Go to item (ii) and repeat the procedure, updating the {\it step counter}, $k$.}
\end{enumerate}

We briefly address the choice for $\Delta c$, as mentioned in item (ii) above. Clearly, $\Delta c$ can be chosen even or odd, but  an odd $\Delta c$ always result in a violation of the condition given in Eq. \ref{EQN-CONSTR2}. On the other hand, according to [Ans08], if $\Delta c$ is taken even, violation does not always happen, hence mimicking the sandpile model (whose avalanche sometimes happen, and sometimes not). Furthermore, they found that an odd $\Delta c$ did not lead to a SOC behavior, whereas an even  $\Delta c$ did. The above scheme produced a SOC behavior so that the average TSN color increased as a function of steps, although in a highly oscillatory manner, probably due to the size of the TSN used, as pointed out by [Che08]. The latter authors revisited the matter using a different prescription, in which $\Delta c$ was not fixed, but randomly chosen between $2$ and a maximum color value (even integers), ranging from values in the set: $c_{\rm max} = \{10, 30, 100, 1000, 10000 \}$. Regarding the choice of initial colors, $\mathsf{C}_{\rm init}$ (item (i) above), [Che08] also differed from [Ans08], as they randomly selected $\mathsf{C}_{\rm init}$ from integers between $1$ and $c_{\rm max}$. However, they did not find a clear effect regarding the choice of $c_{\rm max}$  on the SOC behavior (see [Che08] for details). These authors also used a larger number of vertices, with a probability scheme for color propagation. They found that the average color increased linearly after an initial relaxation period.

\subsection{Self-organized criticality.}

In the late 80s, a new unifying theory explaining the nature of power-law relations arising in complex systems was proposed by Bak et al.\cite{Bak87}, \cite{Bak88}, which subsequently received considerable attention (e.g., \cite{Dha90}, \cite{Marko14}, \cite{Wat16}, and references therein). The basic features of this theory, named {\it self-organized criticality} (SOC), is best understood in terms of the simple dynamics of sandpiles (the first model exemplifying how a system moves to its critical state without any fine-tuning of an external parameter). 

Generally, from an arbitrary initial condition (e.g., a sandpile) and a stochastic evolution, the system reaches a critical state after a sufficient long time, characterized by power-law correlations in space and time. This can be seen in the sandpile model by dropping a grain of sand at arbitrary times and locations on the sandpile, so that an amount of grains of sand may topple down according to certain critical value at each site -- creating a finite avalanche, until equilibrium is again reached. 

For the SOC phenomena, a power-law probability distribution of {\it avalanche size} ($s$) is expected (also known as the {\it Pareto distribution function}, c.f. \cite{Marko14}): 

\begin{equation}
\mathcal {P}(s) = s^{\alpha}, ~~~~~ \alpha < 0.\label{EQ-Pareto}
\end{equation}
\noindent

There are theoretical open questions and new interpretations, specially resulting from the considerable advancement of this field (and from highly heterogeneous applications), which can be found, e.g.,  in the critical review of Ref. \cite{Wat16}. However, there are generally accepted typical features shared by SOC systems, such as a slow external driving perturbation and fast relaxation, so that the driving force does not disturb the running of the avalanche, while it is occurring. Indeed, as it is clear from the propagation rules presented in Sec. \ref{Sec-TSN}, the TSN is randomly disturbed only after it regains a gauge-invariant state. Hence, in the present case, by design, relaxation triggered by a color avalanche is always faster than the interval between external driving perturbations. This feature is preserved in our new models, to be described in the next section.

\section{New propagation rules with a growth scheme for the frozen TSN model \label{Sec-NewTSN}}

The motivation to explore a growth scheme comes from the observation in [Ans08] and [Che08] that the dual space tends to expand due to an overall increase of the mean color of the TSN, as a function of the number of steps. The question we address here is whether expanding models at higher rates than those reported previously still produce SOC behavior. In order to achieve this effect, we designed our models by keeping a clear distinction between the time scales of the external driving perturbation and the internal redistribution towards gauge invariance, but altering the external driving force in a significant manner. That is, in terms of devising an amplitude coupled to a growth model (for an overview of extensions to the sandpile model, see e.g., \cite{Marko14}). Note that such a design did not guarantee SOC behavior beforehand, being only verified {\it a posteriori}. In this section, we describe our models in detail.

\subsection{New models.}

We adopt of a growth scheme for the color disturbance $\Delta c$, by redefining it through the addition of some fixed or variable quantity $\mathcal{C}$, or of a related function of the {\it step counter}, $k$ ($k = 1, \dots, N_{\rm steps}$). Hence, only the steps (ii) and (iv) in the propagation rules (c.f. Sec. \ref{Sec-TSN}) are affected by this scheme. All models start with an initial $\Delta c_{k=1} = 2$. We describe our models as follows.

\subsubsection{Model A:} 

This model represents a simple extension in which a fixed incremental color $\mathcal{C}_{\rm A}$ is added at each step counter $k$ to the color disturbance $\Delta c$. The color addition scheme is given by: 
\begin{equation}
\Delta c_{k+1} = \Delta c_k + \mathcal{C}_{\rm A}, \label{ModA}
\end{equation}
\noindent were we analysed the set $\mathcal{C}_{\rm A} = \{ 2, 20, 200 \}$ (for each run). For example, for $\mathcal{C}_{\rm A} = 2$: $\Delta c_{k=1} = 2, \Delta c_{k=2} = 4, \Delta c_{k=3} = 6, \dots $, etc; for $\mathcal{C}_{\rm A} = 20$: $\Delta c_{k=1} = 2, \Delta c_{k=2} = 22, \Delta c_{k=3} = 42, \dots $, etc..

\subsubsection{Model B:} 

This model modifies the random scheme of [Che08], with a color addition given by: 
\begin{equation}
\Delta c_{k+1} = \Delta c_k + {\rm Rand}(2,\mathcal{C}_{\rm B}),\label{ModB}
\end{equation}
\noindent with $\mathcal{C}_{\rm B} = \{ 4, 20, 200 \}$. ${\rm Rand}$ is a random generator of even integers in the indicated interval (inclusive), updated at each step. For instance, for $\mathcal{C}_{\rm B} =  20$, we could have $\Delta c_{k=1} = 2, \Delta c_{k=2} = 14 ~({\rm for~Rand} = 12), \Delta c_{k=3} =22 ~({\rm for~Rand} = 8), \dots $, etc.

\subsubsection{Model C:} 

This model uses a conditional exponential increase of color addition, as a function of the activation step $k$. The color addition scheme is given by: 
\begin{equation}
\Delta c_{k+1} = \Delta c_k + \mathcal{C}_{\rm C}[f(k)], \label{ModC}
\end{equation}
\noindent where 
\begin{equation}
\mathcal{C}_{\rm C}[f(k)]  =    
\left \{ \begin{array}{rl}
\{2, 20, 200\}, & \mbox{if} ~ f(k) \equiv \exp (10 k / N_{\rm steps}) > \Delta c_k; \\
0,             & \mbox{otherwise.}   
\end{array} 
\right. \label{ModCfk}
\end{equation}
\noindent Note that this model incorporates an exponential behavior modulated by a step function, mimicking the effect of some critical condition on the driving perturbation, which is activated as function of an external time. A few initial tests were performed for choosing the factor $10/N_{\rm steps}$ in the exponential function in Eq. \ref{ModCfk}, in order to avoid extremely noisy results. 

\subsubsection{Model Ans08:} 

We also reproduce the results by [Ans08] \cite{Ans08}. For this model, we ran a set of simulations using different numbers of steps and a somewhat larger number of vertices in the TSN, but with their same propagation rules. These simulations are useful as a reference for our code implementation and for comparison purposes.

\subsection{A note on dynamics and time. \label{SEC-TIME}}

Before describing our simulations, we briefly digress to qualitatively motivate our growth scheme of perturbations. The problem of time in the context of canonical quantization of general relativity is a difficult one \cite{Kuc11}. In the present case, the rules described in the previous section lead to the evolution the TSN from a gauge invariant state to another gauge invariant state, considering the intermediary propagation of edge colors from the random vertices disturbances.  Hence, the steps counter, $k$, inherent to the process of vertex activation and color propagation, can be regarded as ``time-keeping'' device, i.e., an external variable parameterizing the agency of sequential, stochastic disturbances on vertices and their subsequent excitations. 

There is a sense in which those state-to-state changes could be associated with a (at least, partial) order process, akin to a ``temporal'' progression, if we consider that the externally induced disturbances are driven from an orderly (i.e., a sequentially labelable) process arising from a external reservoir. 
Although we do not attribute any physical specification for such a reservoir, in more realistic scenarios in which the TSN can be enlarged or be part of an ensemble of interacting TSNs, it is reasonable to assume that such a reservoir represents a mesoscopic or,  alternatively, a semi-classical spacetime region, from which it can be affected and affect the TSNs. 
Hence, one may consider that the externally induced disturbances may change not only at random intervals but also in scale, for instance, as a backreaction mechanism. In this sense, perturbations of the TSN and the background reservoir could be correlated with a hybrid scheme \cite{Boj21}, a line of investigation which is being currently studied, but which is outside the scope of the present paper.

\subsection{Simulations.} 

\begin{table}
\centering
\caption{Summary of the simulations} \label{TabSim}
\begin{tabular}{|l|l|l|l|l|} \hline
Model                                     & 
$N_{\rm v}$                               & 
$N_{\rm steps}$                           & 
Color Increments ($\mathcal{C}_{\rm A}$,$\mathcal{C}_{\rm B}$ or $\mathcal{C}_{\rm C}$)   & 
$N_{\rm RUNS}$                            \\ \hline
A  ($\Delta c_{k+1} = \Delta c_k + \mathcal{C}_{\rm A}$)              & 
$484$                                     & 
$10^5$                                    & 
$\{ 2,20,200 \}$                          & 
$\{ 5, 5, 5 \}$                           \\
 &  & $10^6$  & $\{ 2,20,200 \}$ & $\{ 5, 5, 5 \}$ \\ 
 &  & $10^7$  & $\{ 2,20,200 \}$ & $\{ 1, 1, 1 \}$ \\ 
                                                                      & 
$3136$                                    & 
$10^5$                                    & 
$\{ 2,200 \}$                             & 
$\{ 1, 1 \}$                              \\
                                                                      & 
$    $                                    & 
$10^7$                                    & 
$\{ 2,200 \}$                             & 
$\{ 1, 1 \}$                              \\
\hline
B  ($\Delta c_{k+1} = \Delta c_k + {\rm Rand}(2,\mathcal{C}_{\rm B})$) & 
$484$                                     & 
$10^5$                                    & 
$ \{ 4,20,200 \}$  & 
$\{ 5, 5, 5 \}$                           \\ 
 &  & $10^6$  & $\{4, 20, 200 \}$ & $\{ 5, 5, 5 \}$  \\  
 &  & $10^7$  & $\{4, 20, 200 \}$ & $\{ 1, 1, 1 \}$  \\
                                                                      & 
$3136$                                    & 
$10^5$                                    & 
$\{4,200 \}$          &   
$\{ 1, 1 \}$                              \\
                                                                      & 
$    $                                    & 
$10^7$                                    & 
$\{4,200 \}$          & 
$\{ 1, 1 \}$                              \\
\hline
C  ($\Delta c_{k+1} = \Delta c_k + \mathcal{C}_{\rm C}[f(k)]$)         & 
$484$                                     & 
$10^5$                                    & 
$\{ 2,20,200 \}$                          & 
$\{ 5, 5, 5 \}$ \\
 &  & $10^6$   & $\{ 2,20,200 \}$ & $\{ 5, 5, 5 \}$ \\ 
 &  & $10^7$   & $\{ 2,20,200 \}$ & $\{ 1, 1, 1 \}$ \\ 
                                                                       & 
$3136$                                    & 
$10^5$                                    & 
$\{ 2,200 \}$                             & 
$\{ 1, 1 \}$                              \\
                                                                       & 
$    $                                    & 
$10^7$                                    & 
$\{ 2,200 \}$                             & 
$\{ 1, 1 \}$                              \\
\hline 
Ans08 ($\Delta c_k = 2$; fixed)                                        & 
$484$                                     & 
$10^5$                                    & 
$0$                                       & 
$5$                                       \\
 &  & $10^6$   & $0$    & $5$ \\
 &  & $10^7$   & $0$    & $1$ \\ 
                                                                       & 
$3136$                                    & 
$10^5$                                    & 
$0$                                       & 
$1$                                       \\
                                                                       & 
$3136$                                    & 
$10^7$                                    & 
$0$                                       & 
$1$                                       \\
 \hline      
\end{tabular}
\end{table}

We developed a {\tt Python} \cite{Python} code (with the {\tt SciPy} \cite{SciPy}  package for curve fitting) to implement the algorithms related to the TSN construction, edge color attribution, gauge-invariant tests, propagation rules (the simulations {\it per se}) and analysis of the results.
Our set of simulations incorporate the new propagation rules given by the models, A, B and C, described in the previous section, and also the Ans08 Model, giving a total of $124$ simulations. The list of simulations and general parameters are summarized in Table \ref{TabSim}, with a few observations, as follows:

\begin{itemize}
\item{The size of the TSN, given by the total number of vertices ($N_{\rm v}$): our simulations were run for both the options $N_{\rm v} = \{ 484, 3136 \}$. Note: [Ans08] used $N_{\rm v} = 361$, whereas [Che08] used $N_{\rm v} = 10000$ (they also explored a range of $N_{\rm v}$ values as a check of the results found in [Ans08]).}
\item{The number of activation steps, chosen from the options $N_{\rm steps} = \{ 10^5, 10^6, 10^7 \}$.}
\item{The color increments, as specified in the previous section. We use the notation $\mathcal{C} \equiv \{ \mathcal{C}_{\rm A}, \mathcal{C}_{\rm B}, \mathcal{C}_{\rm C} \}$, as a proxy for any of the color increments used in these models. Hence, the minimum color increment used was $\mathcal{C} = 2$, and the maximum was $\mathcal{C} = 200$. These color increments result in cumulatively large values of the color perturbation $\Delta c$ as the number of steps is increased, as will be discussed in the next section. Note that $\mathcal{C} = 0$, for the [Ans08] model, as the color perturbation was fixed to $\Delta c_k = 2$ (i.e., no color increment scheme was imposed).}
\item{Number of runs ($N_{\rm RUNS}$) of a given type of simulation, i.e., using the same parameters, but starting from a different (new realization) of initial conditions on the edge colors as well as from a new random selection of activation vertices. Note that initial colors are randomly chosen from integers in the set $\mathsf{C}_{\rm init} = \{10, \dots, 30 \}$ (even), as in [Ans08], under the condition of a global gauge-invariance on the TSN.}
\end{itemize}

Note that we fixed $\mathsf{C}_{\rm init}$ exactly as in [Ans08], even though we could have chosen a different initial set, e.g., including odd integers and different values, as in [Che08]. As mentioned previously (c.f. the end of Sec. \ref{Sec-TSN-1}), the maximum color for $\mathsf{C}_{\rm init}$ did not seem to cause a noticeable difference in the results found in [Che08]. Therefore, at least in this sense, changing the initial maximum color values to other choices would probably produce results insensitive to such specific choices. For concreteness and reference purposes, we chose to restrict $\mathsf{C}_{\rm init}$ to same values as in [Ans08], allowing for a direct comparison with the new models. However, this is an interesting question in itself, i.e. the degree of sensitivity of the SOC behavior to the initial state and the admission of odd colors. An amplification of the initial color space and an analysis of the impact on the results are left for future work.

We point out that the largest TSN runs ($N_{\rm v} = 3136$, $N_{\rm steps} = 10^7$) were computationally very demanding for our current resources, and under our specific {\tt Python} implementation. Hence, we explored a smaller set of the parameters for these runs, and did not perform additional realizations (for the same parameters), as were done for the smaller TSN runs (c.f. $N_{\rm RUNS}$ column in Tab.\ref{TabSim}).

\section{Results \label{Sec-RES}}

\subsection{Avalanches.}

\begin{figure}
\centering
\includegraphics[trim= 0in 1.2in 0in 0in,clip,width=1.0
\linewidth]{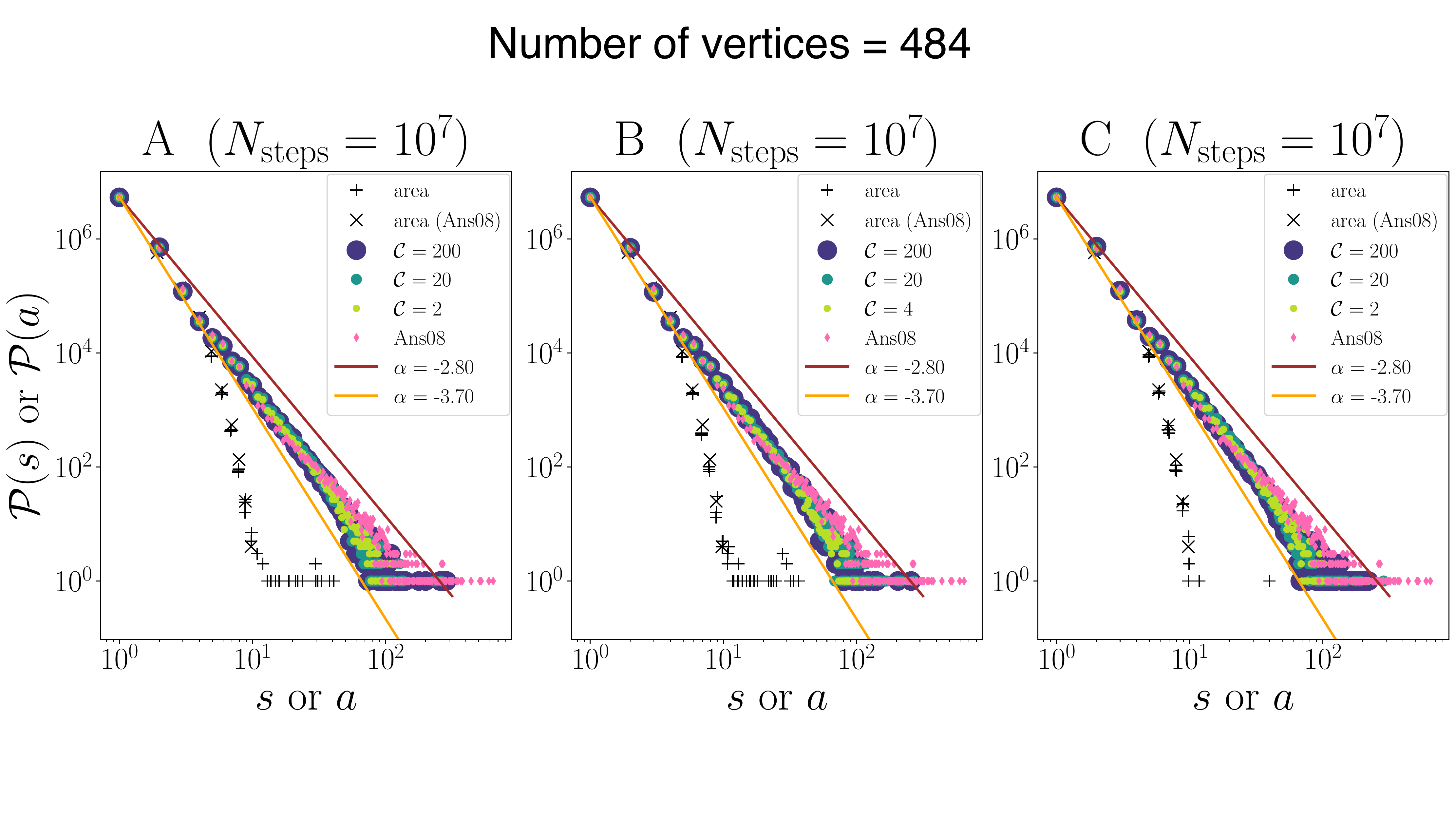} 
\includegraphics[trim= 0in 1in 0in 0in,clip,width=1.0
\linewidth]{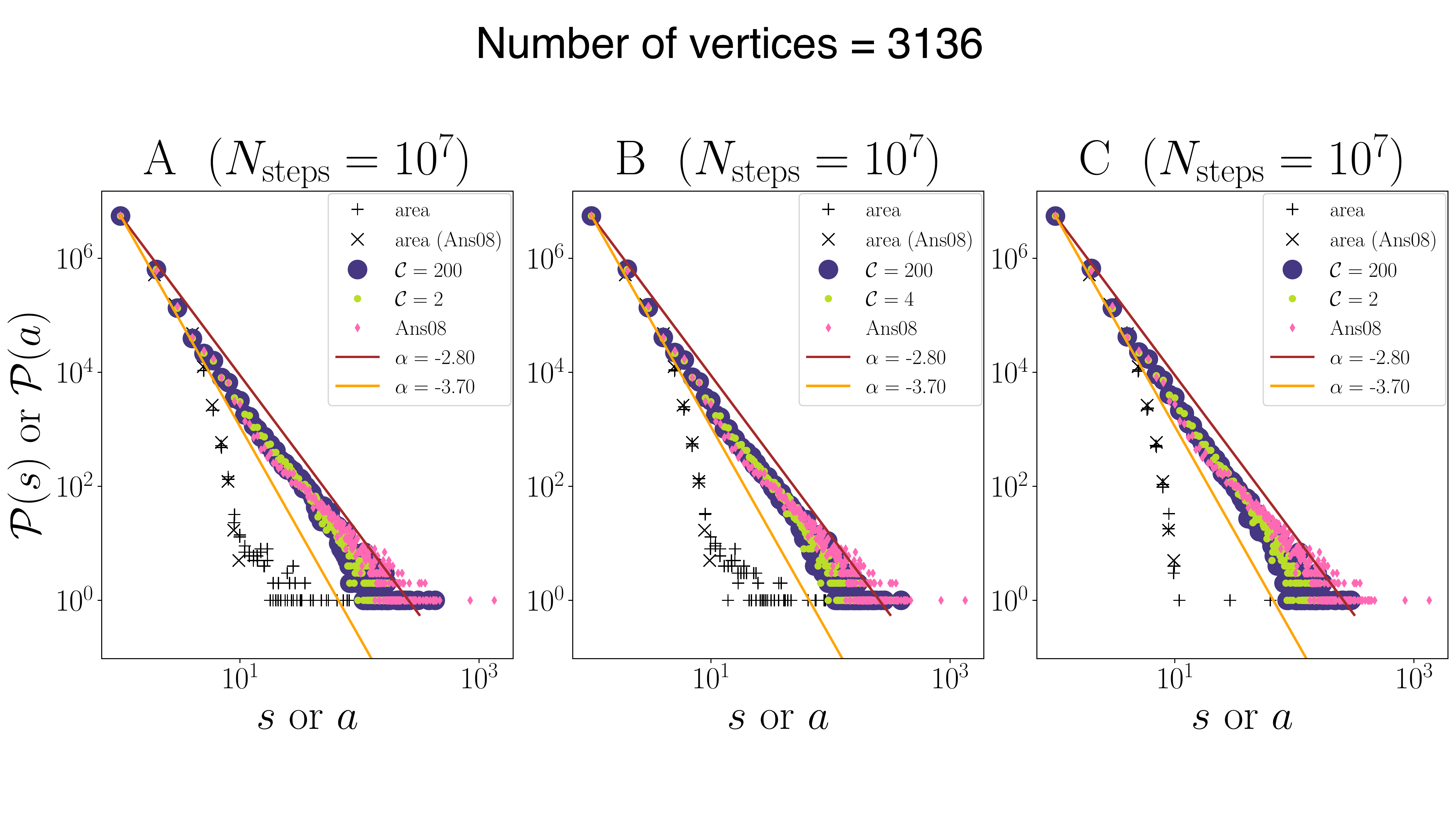} 
\caption{Log-log plots of the distribution of avalanche sizes (circles) and areas (crosses and pluses) for the A, B and C models, with the (single) Ans08 run also displayed in each of these panels for comparison; results are for the $N_{\rm steps} = 10^7$ runs only (c.f. Tab. \ref{TabSim}). Circular symbol sizes are proportional to the color increment $\mathcal{C}$ used. {\it Top:} $N_{\rm v} = 484$; {\it bottom:} $N_{\rm v} = 3136$. Lines ($\alpha = -3.70$ and $\alpha = -2.80$) are for reference only. \label{FIG-AVAL7}}
\end{figure}

We fitted our models to the Pareto distribution, Eq. \ref{EQ-Pareto}; however, we tested for other heavy-tailed distributions, such as the log-normal and log-Cauchy distributions \cite{Marko14}, obtaining worse or sometimes non-convergent fits. The Pareto distribution was adequate in fitting all our models, resulting in different values of the exponent $\alpha$, all within the range $\alpha \in [ -3.23, -2.41 ]$ (to be discussed in more detail in the next subsection). That is, despite the difference in the color increment rules (fixing all the other parameters), all models follow a similar scale invariant behavior, indicative of SOC.

We present the avalanche results of our A, B and C models, for $N_{\rm steps} = 10^7$, in Fig. \ref{FIG-AVAL7}, for TSN sizes of $N_{\rm v} = 484$ (top panel) and $N_{\rm v} = 3136$ (bottom panel), respectively. The corresponding results for the (single) Ans08 run are displayed in the same graphs for comparison.
In order to avoid excessive clutter in our presentation,  we only display here the results for $N_{\rm steps} = 10^7$, leaving a presentation of the other runs in \ref{App}. The panels of Fig. \ref{FIG-AVAL7} show log-log plots of the probability distribution function of avalanches, in terms of avalanche sizes, $\mathcal{P}(s)$ and areas. We include two lines ($\alpha = -3.70$ and $\alpha = -2.80$) for reference only. Stacked areas are also shown, with a behavior qualitatively also similar to those found in [Ans08] (see their Fig. 5).

We confirm the evidence of SOC in the Ans08 model, here analyzed using larger sizes for the TSN and evolving the disturbances for three ranges of $N_{\rm steps}$  (c.f. Tab.\ref{TabSim}). In [Ans08], the best-fitting parameter has been originally found to be: $\alpha = -3.3$ for $N_{\rm steps} = 10^7$. We confirm a close value, of $\alpha = -3.23$, for the run $N_{\rm v} = 3136$, $N_{\rm steps} = 10^7$ (whereas advancing less steps, i.e., for $N_{\rm steps} = 10^5$, we found $\alpha = -3.02$; see next subsection for more details). 

We note in passing that a gauge invariant state was always found for all simulations in the present work. That is, there were no occurrences such as ``infinite'' loops or a stalling of the code. However, it is not clear whether this situation could happen for a very large TSN simulation, and given a large enough number of steps. Indeed, the largest avalanche sizes occurred in the larger TSN runs (see Fig. \ref{FIG-AVAL7}), so it is expected that larger and larger (even though rarer and rarer) avalanches may occur as we increase the TSN. Although the {\it area} of the avalanche is clearly always bounded by the system size, in the limit of an idealized, ``infinitely'' large TSN, it is not clear whether a finite avalanche {\it size} can be always achieved. Indeed, to the knowledge of the present author, this is an open problem  (see, e.g., \cite{Kad89}): how to extrapolate correctly (from a general theory) the infinite-volume properties of a SOC system from its finite-size properties, specially given that most results in the literature are based on simulations (with their inherent numerical limitations). Indeed, in order to estimate uniquely both the scaling function and exponent for a given observable of a SOC system, independently of its size, some {\it Ansatze} have been proposed (c.f. \cite{Marko14} and references therein). However, such an analysis is beyond the scope of the present work.

\subsection{Behavior of the $\alpha$ exponent.}

\begin{figure}
\centering
\includegraphics[width=0.7\linewidth]{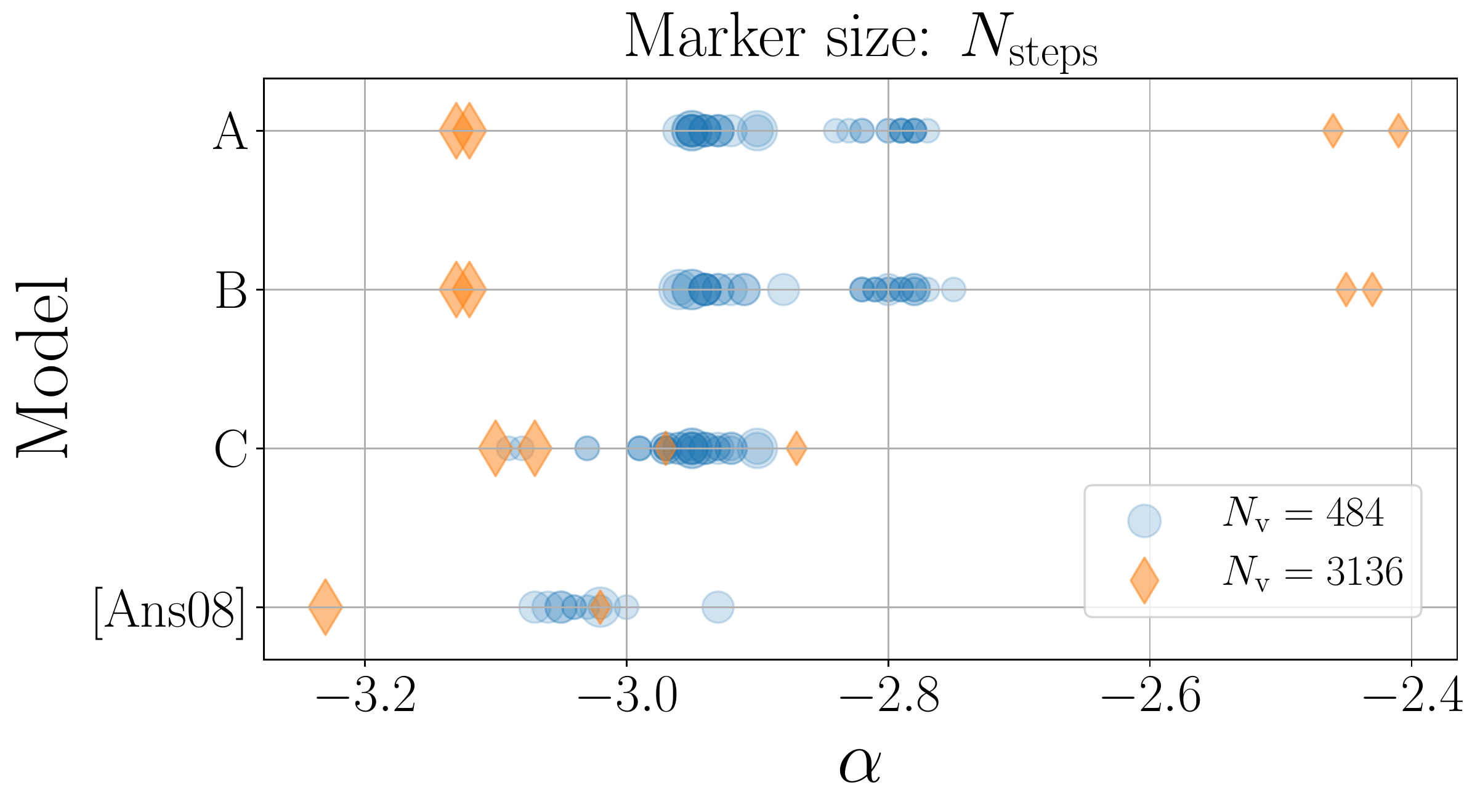} 
\includegraphics[width=0.7\linewidth]{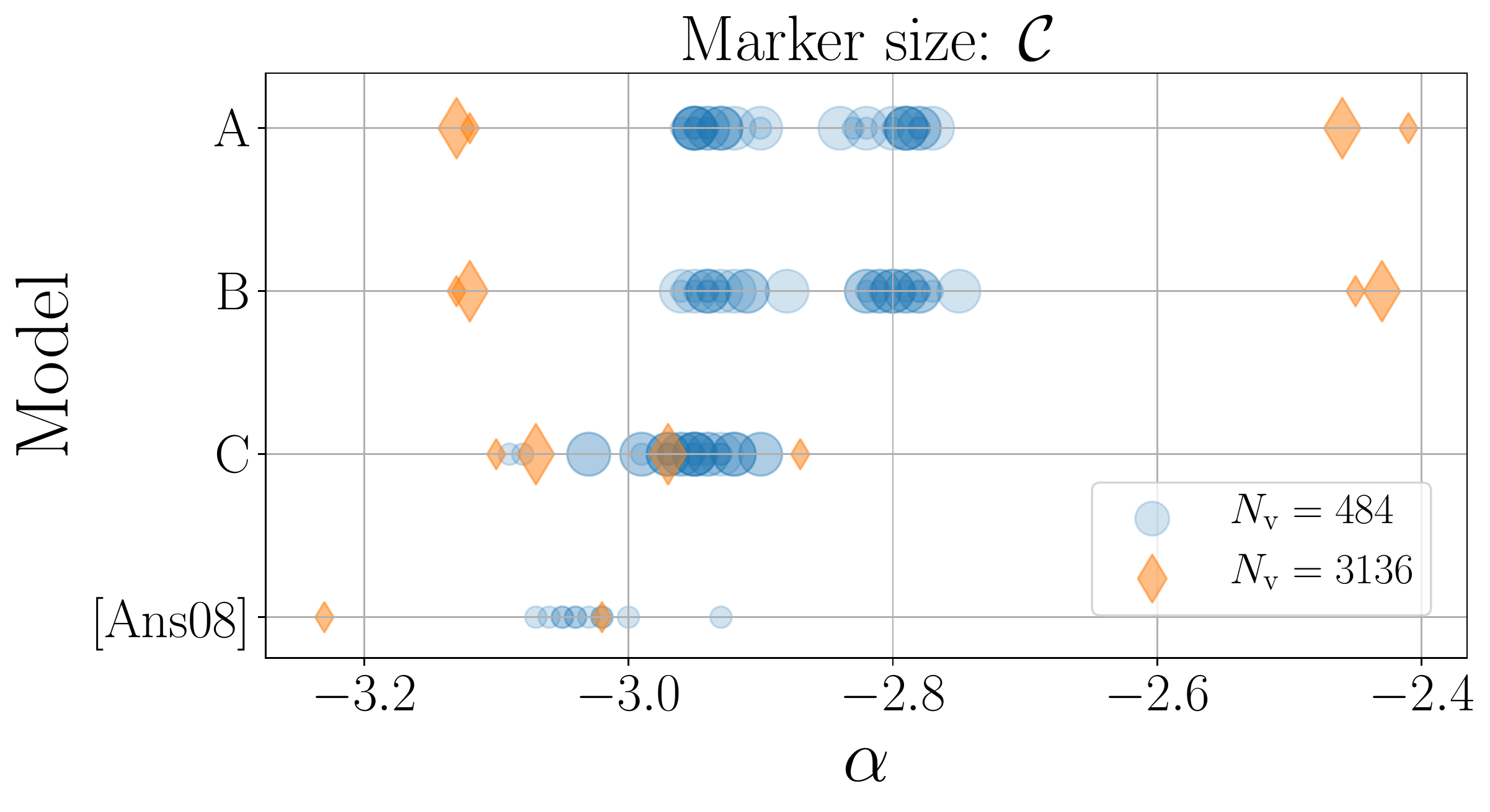} 
\caption{Pareto best-fitting  $\alpha$ values (${P}(s) = s^{\alpha}, \alpha < 0$) for all simulations in this work (c.f. Tab. \ref{TabSim}), as a function of the size of the TSN (circles: $N_{\rm v} = 484$; diamonds: $N_{\rm v} = 3136$). Results are displayed such that the marker (symbol) sizes are proportional to $N_{\rm steps}$ (top panel) and to color increment $\mathcal{C}$ (bottom panel). 
Notice that the exactly same data is being displayed in both panels: the only change is the additional information (signaled by the size of the marker symbol) concerning the dependence of the $\alpha$ values on $N_{\rm steps}$ and $\mathcal{C}$, respectively. 
\label{FIG-ALP}}
\end{figure}

We present in Fig. \ref{FIG-ALP} a summary of the fittings to the log-log plots of the probability distribution function of avalanches, in terms of avalanche sizes, for all simulations in the present work (c.f. Tab. \ref{TabSim} and  \ref{App}). For each model, the Pareto $\alpha$ values are shown as a function of the size of the TSN, and also as a function of $N_{\rm steps}$ and color increment $\mathcal{C}$ (see the corresponding caption for further details). We note the following general trends:

\begin{itemize}
\item {The overall behavior of the $\alpha$ exponent seems to be highly sensitive to $N_{\rm steps}$ in the case of the larger TSN runs ($N_{\rm v} = 3136$), as compared to the smaller TSN runs  ($N_{\rm v} = 484$). In the former case, fewer number of steps (less evolution) result in significant shallower distributions (specially for the A and B models, giving $\alpha > -2.5$), whereas the latter cases result in larger $|\alpha|$ values. The C models, however, show less sensitivity to $N_{\rm steps}$ as compared to the other models.}
\item {The resulting $\alpha$ exponent seems somewhat insensitive to the color increment used, for any fixed model. Take, for instance, model A, regarding the results for the larger TSN runs ($N_{\rm v}=3136$). In the upper panel, consider the two $\alpha$ values at right ($\alpha \sim -2.5$). Their small sized symbols indicate that these results are for the runs with smaller $N_{\rm steps}$($=10^5$).  Compare these with the two larger diamond symbols at right ($\alpha \sim -3.1$) in the same panel, which are the results for the larger $N_{\rm steps}$ ($=10^7$) runs. By comparing those two sets of points with their counterparts in the lower panel, we see that they correspond to different color increments.}
\item {The A and B models show very similar results. The C and Ans08 models show a slight tendency for larger $|\alpha|$ values than the A and B models, for the smaller TSN runs.}
\end{itemize}

\subsection{Mean color evolution.}

\begin{figure}
\centering
\includegraphics[trim= 0in 0in 1.2in 0in,clip,width=1.0
\linewidth]{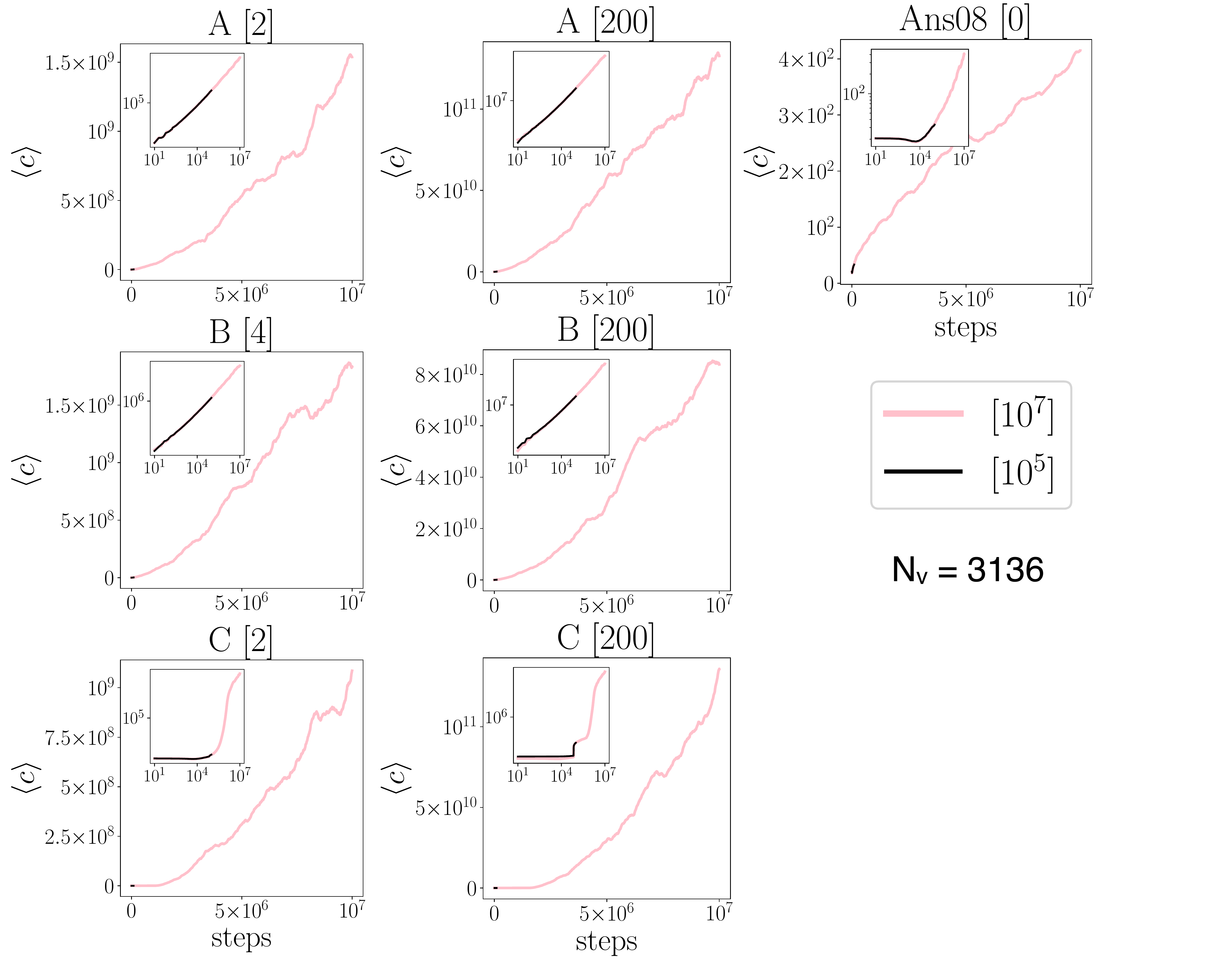} 
\caption{Mean color ($\langle c \rangle$) evolution of the TSN with $N_{\rm v} = 3136$, for each model, and for the color increment $\mathcal{C}$ specified in brackets next to the model label; log-log scale plots are shown in the respective insets. The evolution of $\langle c \rangle$ is shown for independent simulations, each run up to $N_{\rm steps} = 10^5$  and $N_{\rm steps} = 10^7$ (plotted in the same respective panels). \label{FIG-COL57-3136}}
\end{figure}

In Fig. \ref{FIG-COL57-3136}, we present the mean color evolution, $\langle c \rangle$, for the largest TSN ($N_{\rm v} = 3136$), where the mean color is the average over all the edges of the TSN for a
particular step of the dynamics, imposed by the propagation rules of the models. The plots show the evolution up to $N_{\rm steps} = 10^7$, with the results for $N_{\rm steps} = 10^5$ also shown (which were run independently, c.f. Tab. \ref{TabSim}). Insets show the respective evolutions in a log-log scale. 

Given the scales involved, the mean color evolutions of the $N_{\rm steps} = 10^5$ runs closely overlap those at the initial stages of the $N_{\rm steps} = 10^7$ runs,  at the resolution of the graphs. Although they actually statistically disperse, the effect is not significant initially. However, as the evolution proceeds, the outcomes begin to disperse even more. In the \ref{App}, the corresponding figures for the smaller TSN ($N_{\rm v} = 484$) runs give an indication of the statistical dispersion of these curves, c.f. Figs. \ref{FIG-COL56-484} and \ref{FIG-COL567-484}. It is also clear that the mean color evolution of the larger TSN is much less noisy than those of the smaller TSN runs, a result also found by [Che08].

We highlight a few observations on the mean color evolution of the present models:

\begin{itemize}
\item{Clearly, the new models (A, B, C) produce very high mean TSN color values after a certain amount of steps. For models with the smallest color increments ($\mathcal{C}$), the final mean color value reaches $\langle c \rangle _{(10^7)} \sim 10^9$; whereas for the largest color increment, $\langle c \rangle _{(10^7)} \sim 10^{11}$. This is to be contrasted with the much smaller values obtained in the Ans08 model, as expected (i.e., no color increment prescription besides the fixed color perturbation $\Delta c_s = 2$).}
\item{The high values of $\langle c \rangle$ were expected, but the form of those curves were not exactly inferred beforehand. {\it A posteriori}, we  noted that the A and B models behave very similarly, following an approximately straight line in the log-log plots (insets of models A and B in Fig. \ref{FIG-COL57-3136}), which indicates that $\langle c \rangle$ follows a power law relationship with the number of steps. The C models resemble the Ans08 model in the sense that the increase of $\langle c \rangle$ is initially small, but amplifies after $\sim 10^4$ steps. However, the Ans08 model subsequently follows an approximately straight line in the log-log scale plot, whereas the C model increases steeply, and with a more complicated evolution, nevertheless, showing a behavior consistent with the growth scheme for this model. As mentioned previously, however, it was not clear beforehand the exact impact of the growth scheme on the results.}
\end{itemize}

\section{Summary and conclusions \label{Sec-SUMM}}

This work revisited frozen TSN simulations obeying certain evolution rules, as part of an ongoing investigation regarding the notion of criticality in spin networks. It has been suggested \cite{Mar97} that one possible mechanism for the emergence of classical spacetime from a quantum geometric background could be associated with an underlying self-organized critical (SOC) process occurring in the quantum spacetime. At this stage, our aim was to expand previous work \cite{Bor99}, [Ans08], [Che08], by providing new propagation rules as models for perturbations in the TSN, which diverged somewhat from the sandpile model. That is, we investigated the possibility of a stronger and systematic external driving force, as a backreaction model. Our new models closely generated power-law distributed avalanche sizes (SOC), while showing a significantly pronounced increase of the mean color of the TSN with time (here understood as an external variable parameterizing the agency of sequential, stochastic disturbances). The resulting expanding, $2$-dimensional dual spaces showed two basic classes of evolution: one with power-law correlations in space and time, and the other with ``loitering'' and exponential phases. Our work expanded the range of models in which critical states in the TSN could lead to expansion effects in the dual space, without fine-tuning. In the following, we discuss our results in terms limitations, open questions and prospects.

Our simulations are based on finite-size systems, and the results gave truncated power-laws, at the level allowed by the numerical runs. Therefore, our (and previous) results are, more precisely, indicative of SOC. It seems reasonable to expect that the SOC behavior will be persistent in larger and larger simulations, given the fixed connectivity of the frozen TSN, which can be extended arbitrarily with the same local structure. However, extremely large avalanches could potentially take prohibitive large computational times to reach gauge invariance in a simulation setting. In order to infer the ``infinite''-size limit behavior of such models, the development of complementary methods would be necessary (e.g., \cite{Kad89}). Another point is that, for concreteness, we used only even integers for the initial color state of the TSN. A future extension of the parameter space would complement the conditions leading to SOC states in spin networks. 

Our paper did not address why TSN models (present and previous ones) showed indications of SOC behavior (for certain propagation rules and parameters).  A theoretical framework able to connect the specifications of the parameter space of a spin network, at one hand, and the appearance of SOC behavior, at the other hand, is missing. But, to our best knowledge, this involves open questions in the field of dynamical systems itself, at least from a general theoretical standpoint. The lack of a general theory is mostly explainable by the fact that the traditional methods of statistical mechanics are not suitable to SOC systems (e.g., \cite{Ces04}). There are certainly many proposed models that attempt to explain heavy-tailed distributions arising in many natural phenomena (see, e.g. \cite{Marko14} and references therein).  It would be valuable to understand the exact conditions in which the probability of observing extremely large spin avalanches is more likely than that expected from a corresponding exponential distribution.  A possible approach could be to identify and classify results into certain classes \cite{Kad89}, in order to verify how robust the presence of SOC turns out to be, as well as exploring general conditions for SOC failure. Systematic, numerical studies, specifically designed to address the occurrence and failure of SOC in TSNs, and the meaning of this behavior in the context of LQG, would be valuable to advance this matter.

We note, nevertheless, that the available results already point to the existence of a broad pattern for SOC in TSN models: the {\it necessary condition} of a slow driving force (as compared to the timescale of internal rearrangement of the gauge-invariant states), which puts those models in the first class of the three general classes of models for SOC systems (see \cite{Marko14} for a detailed discussion). On the other hand, we are already able to establish that such a condition is {\it not sufficient} to produce SOC, as there are examples of random edge models and random vertices models, both with slow driving perturbations, that do not produce SOC (see examples discussed in [Ans08] and [Che08]). It would be interesting to test whether the breaking of distinct (internal, external) timescales in those models lead to a failure of SOC. 

Other open questions concern a detailed, formal connection of the present line of exploration, in one hand, with the understanding of fundamental problems in LQG, on the other hand, such as the full dynamics and the problem of obtaining the macroscopic limit of the theory (see, e.g., \cite{Dit14}, \cite{Ori14}). These are extraordinarily difficult questions, and outside the scope of the present work. More specifically, our motivation was limited to a more modest goal of exploring the hypothesis of such a connection, rather than providing a formal proof of it.  Our line of research considers spin networks from the standpoint of dynamical systems, which are analysed by the outcome of perturbations in the spin state, considering only fundamental features like the gauge invariance requirement. 
It is an open question of whether spontaneously occurring critical spin states are a natural or a prevalent phenomenon in LQG and, if so, how to connect this mechanism with schemes for constructing the macroscopic limit of LQG. A detailed connection between simulations (designed to test criticality in spin states) and numerical results and techniques in LQG (e.g., \cite{Li19}, \cite{Mie19}, and \cite{Sin12} in the context of loop quantum cosmology) requires further work.

We believe that there is a vast open field relating quantum gravity with the formal theory of dynamical systems, which remains to be explored. For instance, a possible line of research could be the analysis of spin networks modelled as non-equilibrium systems, with an stationary state induced by an external flux of disturbances, with time scales longer than internal time scales. A possible approach can be found in the study of SOC in the context of the Thermodynamic Formalism (TF, c.f. \cite{Rue04}, \cite{Ces04} and references therein), leading to a statistical mechanics extension of SOC systems. This is a promising framework that may provide a formal connection between the probability distribution of avalanches (a characteristics of the critical state) with the microscopic dynamics of a given LQG model. It could also open the possibility of exploring proposed coarse-graining schemes in LQG by translating them into the language of the TF. This possibility is currently being investigated.

An important point of consideration is the following. General relativity in $(2+1)$-dimensions presents fundamental differences with respect to the theory in $(3+1)$-dimensions, which impact quantization schemes accordingly \cite{Car05}. For instance, when considering the $3$-dimensional Ponzano-Regge model (e.g., \cite{RovVidBOOK20} and references therein), the topological invariance of the theory renders the associated quantum theory insensitive to refinements of the discretization scheme. Despite such drawbacks, reduced quantum gravity models have been proved relevant as toy models for the full theory, a feature which might be also true for lower dimensional simulations of criticality in TSNs. For instance, it would be interesting to construct a connection between the asymptotic (stationary SOC regime of the TSN) and the large spin limit relation of the Ponzano-Regge model, as a preliminary step towards a connection between both descriptions.

In terms of applications, we believe that present-day problems in quantum cosmology could be addressed in a complementary manner from the standpoint of criticality but, given the incipience of this field, many questions are open. For example  (\cite{Boj15},  \cite{Mie14} and references therein), is criticality connected to the nature of signature changes, and/or to phase-transitions in the gravitational sector in the early Universe? A link with hybrid schemes in quantum cosmology \cite{Boj21} would also be very interesting. Another potentially fruitful connection could arise from recent numerical results \cite{GozVid21}, in which large variances in geometric variables were found in a LQG computation, with stochastic correlations that suggest an alternative model to inflation. It would be valuable to address the behavior of this class of LQG models in the present context of criticality.

\vspace{1cm}
 
\ack We thank the referees for detailed comments that helped improving the manuscript.  

\vspace{1cm}

\appendix 
\setcounter{section}{0}

\section{Supplementary results for the smaller TSN ($N_{\rm v} =484$) \label{App}}

In this appendix, we present the complete set of graphics related to the distribution of avalanches and mean color evolutions, for the $N_{\rm v} = 484$ runs, which were omitted in the main text for clarity. In the case of avalanche graphs for $N_{\rm steps} = 10^7$, the results for this smaller TSN were already shown in the main text (Fig. \ref{FIG-AVAL7}). Here, we present the supplementary results and briefly discuss them.

\begin{figure}
\centering
\includegraphics[width=1.0\linewidth]{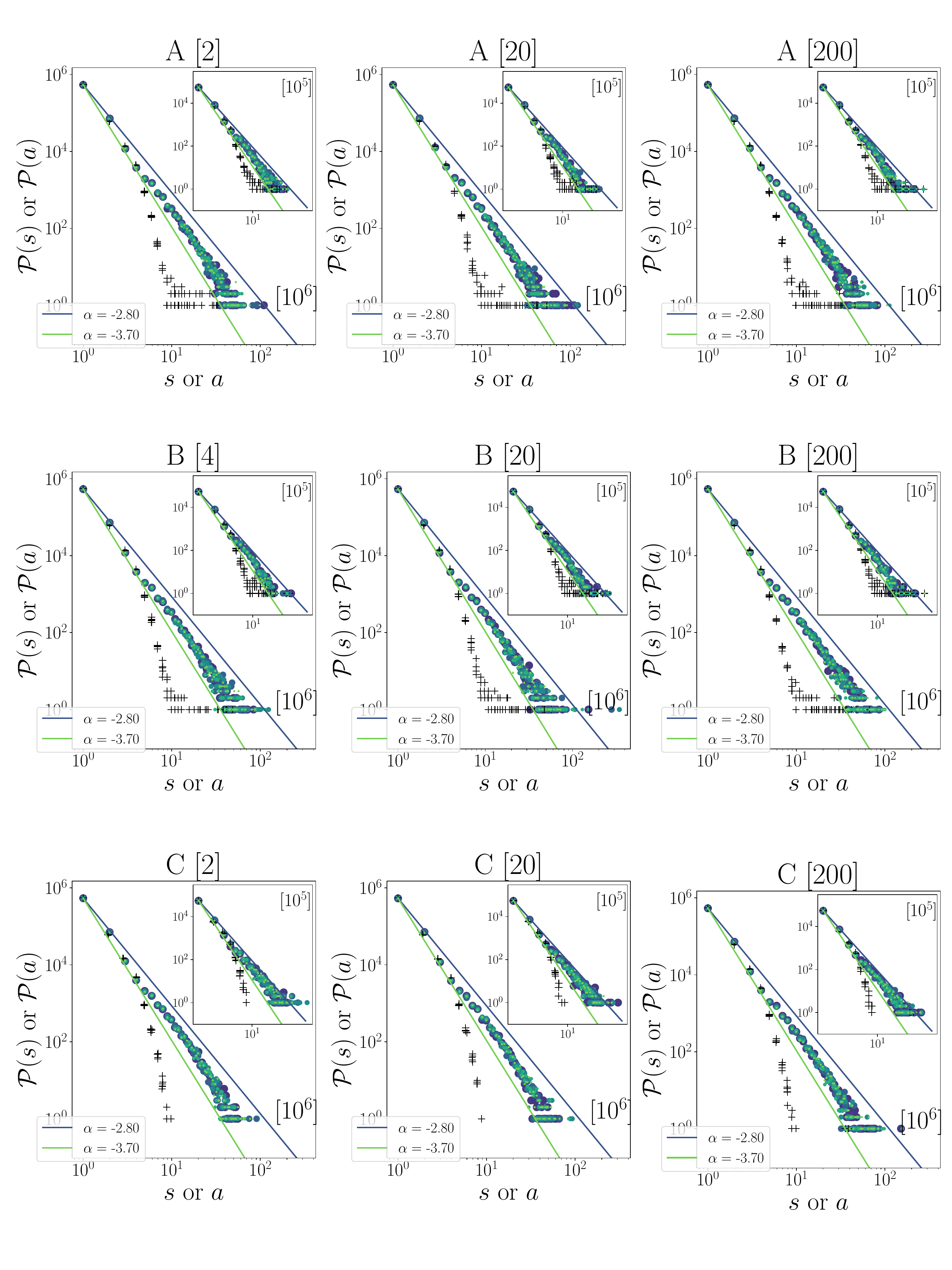} 
\caption{ Log-log plots of the distribution of (stacked) avalanche sizes (circles) and areas (pluses) for the A, B and C models in the smaller TSN ($N_{\rm v} = 484$). Color increment ($\mathcal{C}$) is specified in brackets next to the model label. Results are for the $N_{\rm steps} = 10^6$ (main panels) and $N_{\rm steps} = 10^5$ (insets); c.f. Tab. \ref{TabSim}. Each of these models had $5$ independent runs, stacked in the same respective panels, with different marker colors. Lines ($\alpha = -3.70$ and $\alpha = -2.80$) are for reference only.   \label{FIG-AVAL56-484}}
\end{figure}

\begin{figure}
\centering
\includegraphics[width=0.4\linewidth]{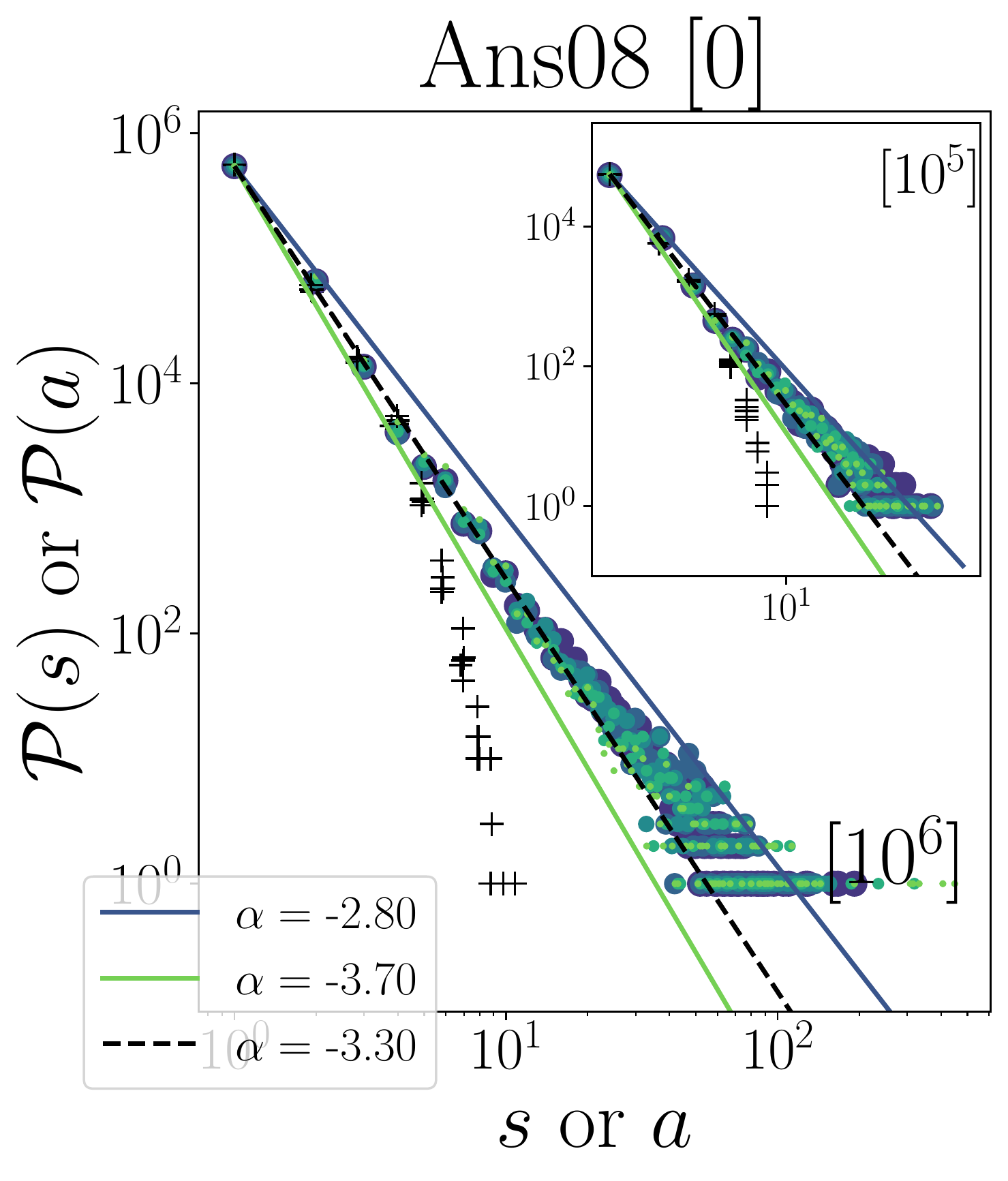} 
\caption{ Same as the previous figure, for the Ans08 model only ($N_{\rm v} = 484$). Black dashed line shows the best-fitting of $\alpha$ found in [Ans08].  \label{FIG-AVAL56-484-Ans08}}
\end{figure}

In the case of avalanches, for $N_{\rm steps} = \{ 10^5, 10^6 \}$, $N_{\rm v} = 484$, Figs. \ref{FIG-AVAL56-484} and \ref{FIG-AVAL56-484-Ans08} show the results for the A, B and C models and for the Ans08 model, respectively. Clearly, all models show evidence for power-law distributed avalanche sizes (SOC), even considering that these models were run on a smaller TSN and for shorter number of steps, as compared to runs using the larger TSN ($N_{\rm v} =3136$), up to $N_{\rm steps} = 10^7$; c.f. Fig. \ref{FIG-AVAL7}.

\begin{figure}
\centering
\includegraphics[trim= 0in 0in 2.3in 0in,clip,width=1.0
\linewidth]{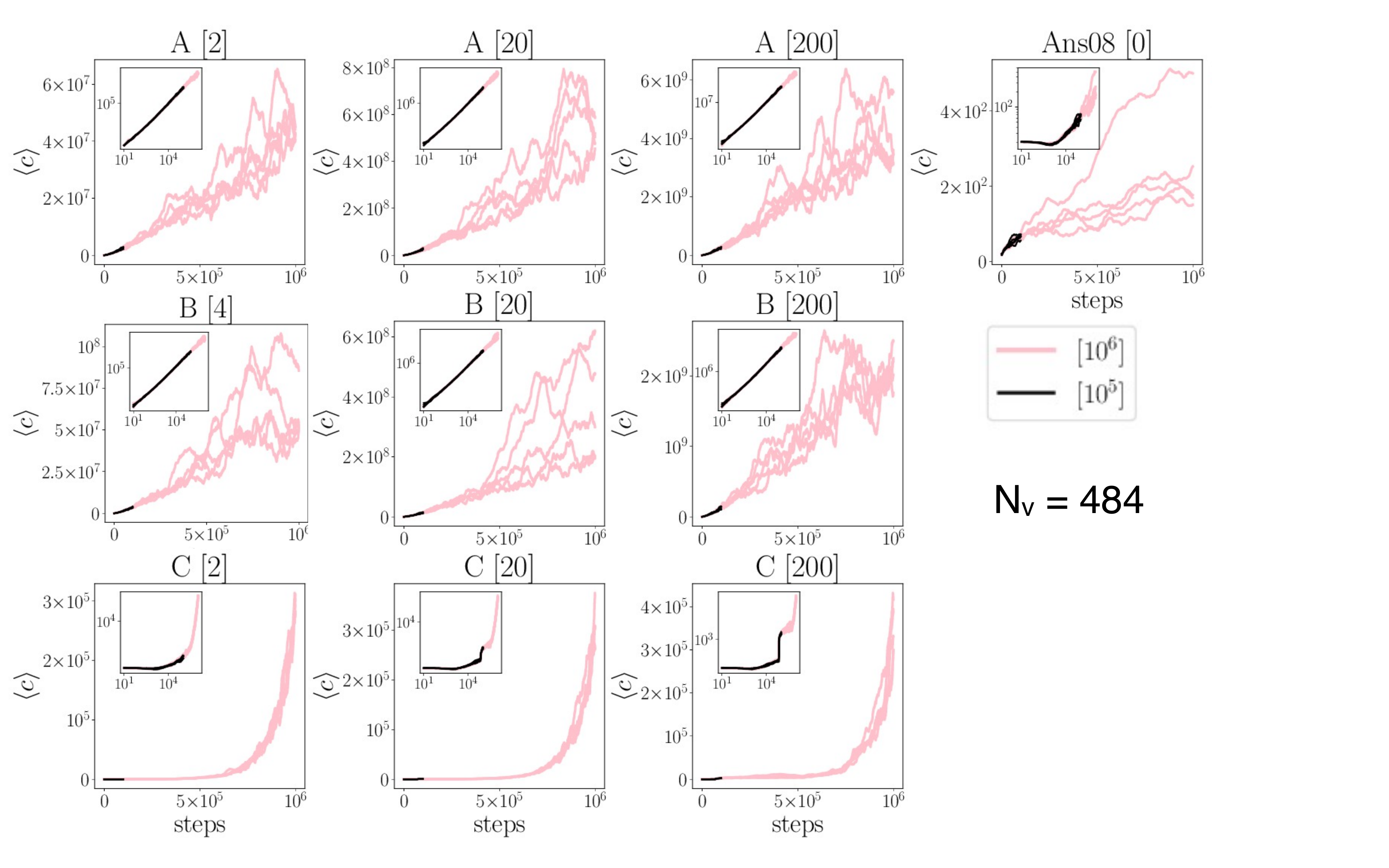} 
\caption{ Mean color ($\langle c \rangle$) evolution of the TSN with $N_{\rm v} = 484$, for each model, and for the color increment $\mathcal{C}$ specified in brackets next to the model label; log-log scale plots are shown in the respective insets. The evolution of $\langle c \rangle$ is shown for independent simulations, each run up to $N_{\rm steps} = 10^5$  and $N_{\rm steps} = 10^6$ (plotted in the same respective panels). \label{FIG-COL56-484}}
\end{figure}

\begin{figure}
\centering
\includegraphics[trim= 0in 0in 1.2in 0in,clip,width=1.0
\linewidth]{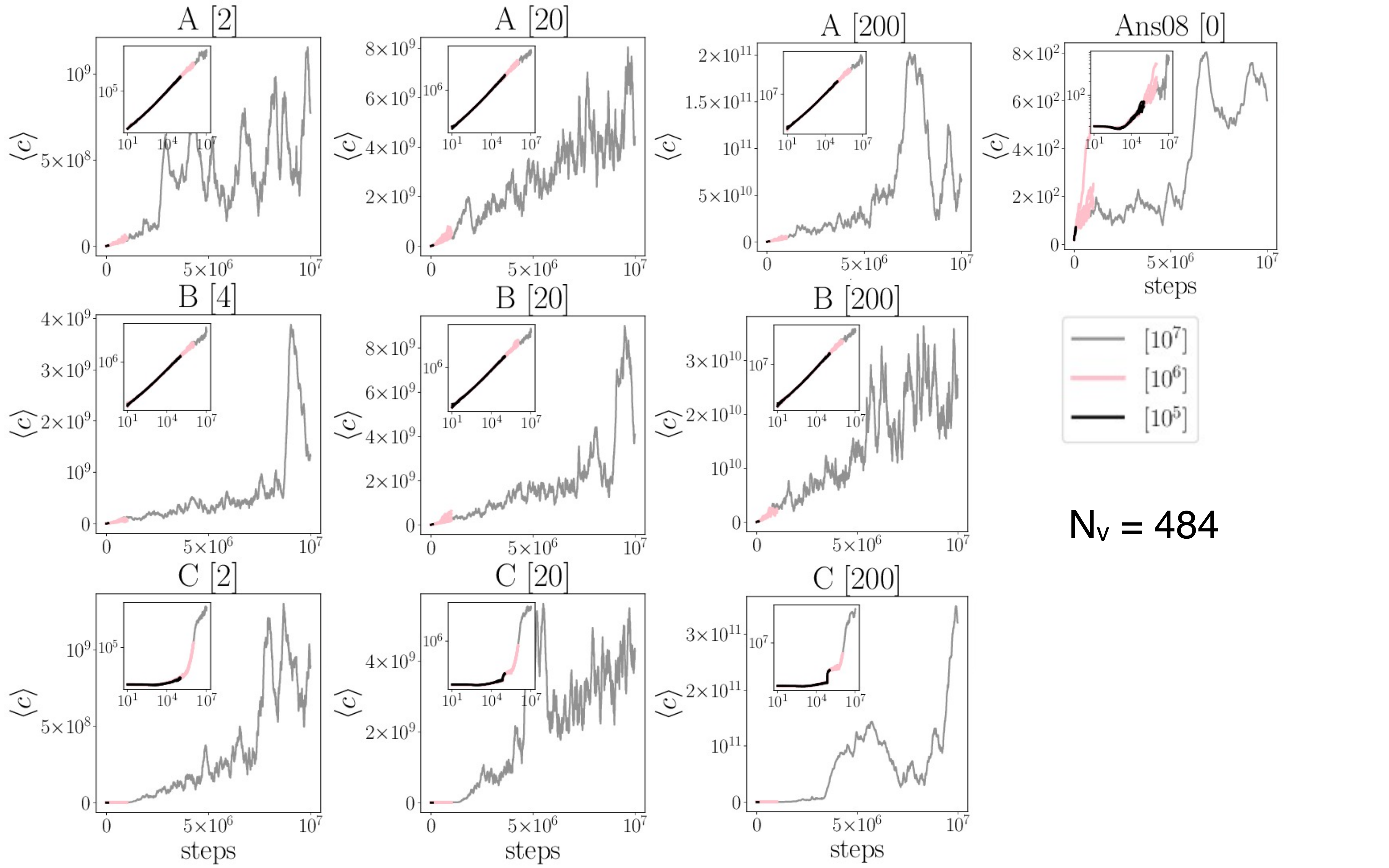}  
\caption{Same as the previous figure, but including simulations run up to $N_{\rm steps} = 10^7$.  \label{FIG-COL567-484}}
\end{figure}

In Figs. \ref{FIG-COL56-484} and \ref{FIG-COL567-484}, we present the mean color evolution of the TSN, for different time scales, respectively, for up to $N_{\rm steps} = 10^6$, and up to $N_{\rm steps} = 10^7$, in order to better visualize the graphics in these different phases.  Note that these cases were run independently, forming therefore an ensemble of different realizations of the same models, from different initial conditions, and evolving up to the corresponding $N_{\rm steps}$ (c.f. Tab \ref{TabSim}).

As mentioned in the main text, the results for the smaller TSN showed a more noisy $\langle c \rangle$ evolution, than using the TSN with $N_{\rm v} = 3136$ (c.f. Fig. \ref{FIG-COL57-3136}), in agreement with [Che08]. We also noted that, using the smaller TSN, all models, including the Ans08 model, present a considerable amplification of the noise for larger time scales (i.e., steps greater than $\sim 10^6$). The qualitative form of the mean color evolution also presents a more complicated behavior at these larger time scales than in the initial steps.

\clearpage

\section*{References}

\bibliography{CCDANTAS_P1a}

\providecommand{\newblock}{}
\begin{thebibliography}{10}
\expandafter\ifx\csname url\endcsname\relax
  \def\url#1{{\tt #1}}\fi
\expandafter\ifx\csname urlprefix\endcsname\relax\def\urlprefix{URL }\fi
\providecommand{\eprint}[2][]{\url{#2}}

\bibitem{Rov98}
Rovelli C 1998 {\em Living Reviews in Relativity\/} {\bf 1}
  \urlprefix\url{https://doi.org/10.12942/lrr-1998-1}

\bibitem{RovBOOK07}
Rovelli C 2007 {\em Quantum Gravity\/} Cambridge Monographs on Mathematical
  Physics (Cambridge University Press) ISBN 978-0521715966

\bibitem{ThiBOOK08}
Thiemann T 2008 {\em Modern Canonical Quantum General Relativity\/} Cambridge
  Monographs on Mathematical Physics (Cambridge University Press) ISBN
  978-0521741873

\bibitem{RovVidBOOK20}
{Rovelli} C and {Vidotto} F 2020 {\em {Covariant Loop Quantum Gravity}\/}
  (Cambridge University Press) ISBN 978-1108810258

\bibitem{Smo05}
{Smolin} L 2005 {\em arXiv e-prints\/} hep-th/0507235 (\textit{Preprint}
  \eprint{hep-th/0507235})

\bibitem{Pen71}
{Penrose} R 1971 {\em Angular momentum: an approach to combinatorial
  space-time\/} (Cambridge University Press) pp 151--180
  \urlprefix\url{https://math.ucr.edu/home/baez/penrose/}

\bibitem{Smo04}
{Smolin} L 2004 {An Invitation to Loop Quantum Gravity} {\em Quantum Theory and
  Symmetries\/} pp 655--682 (\textit{Preprint} \eprint{hep-th/0408048})

\bibitem{Bae00}
{Baez} J~C 2000 {\em An Introduction to Spin Foam Models of BF Theory and
  Quantum Gravity\/} ({\em Lecture Notes in Physics\/} vol 543) (Springer,
  Berlin, Heidelberg) p~25 ISBN 978-3-662-14287-5

\bibitem{Mar97}
{Markopoulou} F and {Smolin} L 1997 {\em Nuclear Physics B\/} {\bf 508}
  409--430 (\textit{Preprint} \eprint{gr-qc/9702025})

\bibitem{Bak87}
Bak P, Tang C and Wiesenfeld K 1987 {\em Phys. Rev. Lett.\/} {\bf 59}(4)
  381--384 \urlprefix\url{https://link.aps.org/doi/10.1103/PhysRevLett.59.381}

\bibitem{Bak88}
Bak P, Tang C and Wiesenfeld K 1988 {\em Phys. Rev. A\/} {\bf 38}(1) 364--374
  \urlprefix\url{https://link.aps.org/doi/10.1103/PhysRevA.38.364}

\bibitem{Dha90}
Dhar D 1990 {\em Phys. Rev. Lett.\/} {\bf 64}(14) 1613--1616
  \urlprefix\url{https://link.aps.org/doi/10.1103/PhysRevLett.64.1613}

\bibitem{Marko14}
{Markovi{\'c}} D and {Gros} C 2014 {\em Phys. Rep.\/} {\bf 536} 41--74
  (\textit{Preprint} \eprint{1310.5527})

\bibitem{Wat16}
{Watkins} N~W, {Pruessner} G, {Chapman} S~C, {Crosby} N~B and {Jensen} H~J 2016
  {\em Space Sci. Rev.\/} {\bf 198} 3--44 (\textit{Preprint}
  \eprint{1504.04991})

\bibitem{Bor99}
{Borissov} R and {Gupta} S 1999 {\em PRD\/} {\bf 60} 024002 (\textit{Preprint}
  \eprint{gr-qc/9810024})

\bibitem{Ans08}
{Ansari} M~H and {Smolin} L 2008 {\em Classical and Quantum Gravity\/} {\bf 25}
  095016 (\textit{Preprint} \eprint{https://arxiv.org/abs/hep-th/0412307})

\bibitem{Che08}
{Chen} J~Z and {Zhu} J~Y 2008 {\em International Journal of Modern Physics A\/}
  {\bf 23} 3891--3899 (\textit{Preprint}
  \eprint{https://arxiv.org/abs/gr-qc/0701175})

\bibitem{Kuc11}
{Kucha{\v{r}}} K~V 2011 {\em International Journal of Modern Physics D\/} {\bf
  20} 3--86

\bibitem{Boj21}
{Bojowald} M and {Ding} D 2021 {\em Journal of Cosmology and Astroparticle
  Physics\/} {\bf 2021} 083 (\textit{Preprint} \eprint{2011.03018})

\bibitem{Python}
Foundation P~S 2019 Python language reference, version 3.7.6
  \urlprefix\url{http://www.python.org}

\bibitem{SciPy}
Virtanen P, Gommers R, Oliphant T~E, Haberland M, Reddy T, Cournapeau D,
  Burovski E, Peterson P, Weckesser W, Bright J, {van der Walt} S~J, Brett M,
  Wilson J, Millman K~J, Mayorov N, Nelson A~R~J, Jones E, Kern R, Larson E,
  Carey C~J, Polat {\.I}, Feng Y, Moore E~W, {VanderPlas} J, Laxalde D,
  Perktold J, Cimrman R, Henriksen I, Quintero E~A, Harris C~R, Archibald A~M,
  Ribeiro A~H, Pedregosa F, {van Mulbregt} P and {SciPy 10 Contributors} 2020
  {\em Nature Methods\/} {\bf 17} 261--272

\bibitem{Kad89}
Kadanoff L~P, Nagel S~R, Wu L and Zhou S~m 1989 {\em Phys. Rev. A\/} {\bf
  39}(12) 6524--6537
  \urlprefix\url{https://link.aps.org/doi/10.1103/PhysRevA.39.6524}

\bibitem{Ces04}
{Cessac} B, {Blanchard} P, {Kr{\"u}ger} T and {Meunier} J~L 2004 {\em Journal
  of Statistical Physics\/} {\bf 115} 1283--1326 (\textit{Preprint}
  \eprint{https://arxiv.org/abs/nlin/0209038})

\bibitem{Dit14}
{Dittrich} B 2014 {\em arXiv e-prints\/} arXiv:1409.1450 (\textit{Preprint}
  \eprint{1409.1450})

\bibitem{Ori14}
{Oriti} D 2014 {\em arXiv e-prints\/} arXiv:1408.7112 (\textit{Preprint}
  \eprint{1408.7112})

\bibitem{Li19}
{Li} K, {Li} Y, {Han} M, {Lu} S, {Zhou} J, {Ruan} D, {Long} G, {Wan} Y, {Lu} D,
  {Zeng} B and {Laflamme} R 2019 {\em Communications Physics\/} {\bf 2} 122
  (\textit{Preprint} \eprint{1712.08711})

\bibitem{Mie19}
{Mielczarek} J 2019 {\em Universe\/} {\bf 5} 179 (\textit{Preprint}
  \eprint{1810.07100})

\bibitem{Sin12}
{Singh} P 2012 {\em Classical and Quantum Gravity\/} {\bf 29} 244002
  (\textit{Preprint} \eprint{1208.5456})

\bibitem{Rue04}
Ruelle D 2004 {\em Thermodynamic Formalism: The Mathematical Structure of
  Equilibrium Statistical Mechanics\/} 2nd ed Cambridge Mathematical Library
  (Cambridge University Press)

\bibitem{Car05}
{Carlip} S 2005 {\em Living Reviews in Relativity\/} {\bf 8} 1
  (\textit{Preprint} \eprint{gr-qc/0409039})

\bibitem{Boj15}
{Bojowald} M and {Mielczarek} J 2015 {\em Journal of Cosmology and
  Astroparticle Physics\/} {\bf 2015} 052 (\textit{Preprint}
  \eprint{1503.09154})

\bibitem{Mie14}
{Mielczarek} J 2014 {\em arXiv e-prints\/} arXiv:1404.0228 (\textit{Preprint}
  \eprint{https://arxiv.org/abs/1404.0228})

\bibitem{GozVid21}
{Gozzini} F and {Vidotto} F 2021 {\em Frontiers in Astronomy and Space
  Sciences\/} {\bf 7} 118

\end{thebibliography}


\end{document}